\documentclass[
reprint,
aps,
amsmath,amssymb,
superscriptaddress
]{revtex4-2}
\RequirePackage{lineno}
\usepackage[symbol]{footmisc}
\usepackage{graphicx} 
\usepackage{dcolumn} 
\usepackage{bm}
\usepackage{pifont}
\usepackage{epsfig}
\usepackage{psfrag}
\usepackage{epstopdf}
\usepackage{wasysym}
\usepackage[usenames]{color}
\usepackage{setspace}
\usepackage{hyperref}
\usepackage{wrapfig}
\usepackage[normalem]{ulem}
\usepackage{xcolor}
\usepackage{lineno}
\usepackage[resetlabels,labeled]{multibib}
\usepackage{verbatim}
\newcites{SI}{References for SI}
\graphicspath{ {./figs-PNAS/}}

\hypersetup{colorlinks = true, citecolor=blue, linkcolor=blue, urlcolor=blue}

\renewcommand{\figurename}{\textbf {Figure}}
\renewcommand{\selectlanguage}[1]{}

\begin{document}
\title{ 
Symmetry Transitions Beyond the Nanoscale in Pressurized Silica Glass
}
\author{Zhen Zhang}
\email[Corresponding author: ]{zhen.zhang@cdut.edu.cn}
\affiliation{College of Physics, Chengdu University of Technology, Chengdu 610059, China}
\author{Zhencheng Xie}
\affiliation{College of Physics, Chengdu University of Technology, Chengdu 610059, China}
\author{Walter Kob}
\email[Corresponding author: ]{walter.kob@umontpellier.fr}
\affiliation{Department of Physics, University of Montpellier, CNRS, F-34095 Montpellier, France}
\date{\today}

\begin{abstract}
Silica is the paradigmatic network glass-former and understanding its response to pressure 
is essential for comprehending the mechanical properties of silica-based materials and the behavior of silicate melts in the Earth's interior. 
While pressure-induced changes in the short-range structure—particularly the breakdown of tetrahedral symmetry—have been well documented
, structural transformations on larger length scales, important for many material properties, remain poorly understood. Here, we numerically investigate the three-dimensional structure of silica glass as a function of compression up to $P \approx 100$~GPa.  
Using a novel many-body correlation function
, we reveal a complex medium-range order: While  for $P \lesssim 10$~GPa, one finds tetrahedral, octahedral, and cubic symmetries, the structure at higher $P$s exhibits alternating cubic and octahedral particle arrangements.
The $P$-dependence of the corresponding structural correlation length displays two distinct maxima, which permits to rationalize the anomalous compressibility of silica.
The identified complex structural organization on intermediate range scales is the result of a pressure-and scale-dependent interplay between directional bonding, packing efficiency, and network stiffness. Since these competing effects are common in network glass-formers, the identified three-dimensional medium-range order, and hence the physical properties of the glass, are expected to be universal features of such materials under extreme conditions.
\end{abstract}

\maketitle
\renewcommand{\thefootnote}


\newpage
Silica is arguably one of the most complex and abundant materials on earth~\cite{heaney2018silica,millot2015silica,hu2015polymorphic,bykova2018metastable,wondraczek2022advancing}. 
Experiments have explored the phase diagram of crystalline silica up to pressures of 271~GPa~\cite{kuwayama2005pyrite}, i.e., conditions found in the outer core of the Earth,
and revealed for pressures above 30~GPa the presence of a multitude of polymorphs, such as coesite-IV and coesite-V which contain [SiO$_5$] polyhedra and face-sharing [SiO$_6$] octahedra~\cite{hu2015polymorphic, bykova2018metastable}, documenting the structural complexity of the material. 
Theoretical studies went up to $P=10$~TPa, pressures 
at which the coordination number of silicon has increased from four to ten~\cite{lyle2015prediction}.

The complex phase behavior of crystalline silica is mirrored in the short-range structure of amorphous silica in that one finds in the latter, as a function of pressure and temperature, local structural motifs and connectivities that are similar to specific crystalline polymorphs~\cite{meade_effect_1988,kono_experimental_2022,sato_sixfold-coordinated_2008,murakami_spectroscopic_2010,sato_high-pressure_2010,brazhkin_atomistic_2011,zeidler_high-pressure_2014,zeidler2014packing,petitgirard_sio_2017, prescher_beyond_2017,murakami_ultrahigh-pressure_2019,hasmy_percolation_2021}.
The nature of the medium range order (MRO) (distances between 5-20~\AA), has, however, remained poorly understood~\cite{ryuo_ab_2017,onodera_structure_2020,kono_experimental_2022}, despite its importance for the dynamical and mechanical properties of glasses and liquids~\cite{stebbins1992structure,horbach1996finite,duval2004origin,leonforte2006inhomogeneous,rouxel_elastic_2007,bykova2018metastable,yu2022silica}.

Indirect insight into the MRO has so far mainly been obtained from two-point correlation functions such as the radial distribution function $g(r)$ or the static structure factor $S(q)$, with $q$ the wave vector, associating position and intensity of the first sharp diffraction peak of $S(q)$ with the characteristic length scale and the orderliness of the structure~\cite{elliott_medium-range_1991,shi2019distinct}. 
However, $S(q)$ does not reveal the nature of the MRO  since the 3D information on the structure is projected onto a 1D quantity.
To overcome this problem, we use here a novel many-body correlation function to unveil how the 3D MRO 
of silica glass depends on $P$. For this, 
we use atomistic simulations with an effective potential~\cite{sundararaman_new_2018,zhang2020potential} to generate realistic silica samples with 570,000 atoms (see Figs.~\ref{SI_Fig1}-\ref{SI_Fig3}) and put them 
under hydrostatic pressure up to 100~GPa 
(see Methods for details). We find that the structural length scales associated with this MRO exhibit a non-monotonic $P$-dependence due to the change of short-range order and the morphology of the Si-O rings. Our findings demonstrate that many-body correlations allow to uncover the complex $P$- and $r$-dependence of the structural order in amorphous silica and other network glass-formers, an important step in advancing our understanding of the mechanical properties of these materials. 

\section*{Results and Discussion}
\begin{figure*}[ht]
\centering
\includegraphics[width=0.95\textwidth]{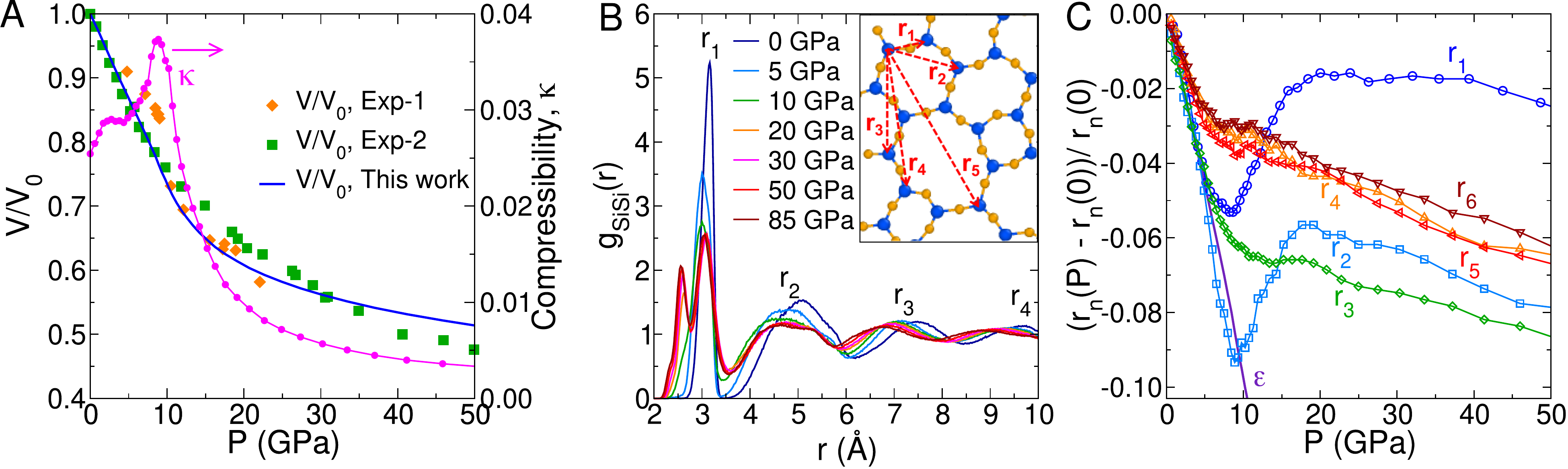}
\caption{Equation of state and pair distribution function. (A) Pressure dependence of volume, normalized by its value at $P=0$ (left ordinate). Experimental data are from Refs.~\cite{sugiura1981dynamic,sato_high-pressure_2010}.
Right ordinate: The compressibility $\kappa$ shows two maxima before it decays monotonically at high $P$. (B) The SiSi pair distribution function $g_{\rm SiSi}(r)$ for different pressures. 
The inset is a two-dimensional schematic diagram of the network structure of silica glass, featuring rings of different size. Typical distances corresponding to the various peak positions are indicated by the arrows.  (C) Relative change of the peak positions of $g_{\rm SiSi}(r)$. $r_1$ to $r_6$ for $P=0$ are respectively 3.17, 5.07, 7.49, 9.65, 11.95, and 14.13~\AA. The solid line is the compressive strain $\epsilon = (L(P)-L(0))/L(0)$ which corresponds to an affine deformation. Figure~\ref{SI_Fig1} shows zooms of panels (A) and (C) on the pressure range 0-20~GPa.
}
\label{Fig1}
\end{figure*}

{\bf Two-point correlation and compressibility.} 
Figure~\ref{Fig1}A (left ordinate) shows that the equation of state agrees well with experimental data~\cite{sugiura1981dynamic,sato_high-pressure_2010} (see Fig.~\ref{SI_Fig1}A for a zoom on the low-$P$ regime). 
Also the simulated structure factor $S(q)$ reproduces the experimental neutron and X-ray data with high accuracy across a wide pressure range (Fig.~\ref{SI_Fig2}), demonstrating that the presented structural models are realistic. 
The compressibility, $\kappa= - V^{-1} dV/dP$ (right ordinate), reaches a weak maximum at around 2.5~GPa, in agreement with the maximum seen in experiments at 2-3~GPa~\cite{bridgman_compressibility_1925}. 
It has been conjectured that this maximum is due to the floppy modes present in a mixture of high- and low-density amorphous domains~\cite{clark_mechanisms_2014}, but below we will show that it can also be related to the change of the MRO. 
Increasing $P$ further induces plastic yielding at around 6~GPa (see Fig.~\ref{SI_Fig3}), just slightly below the corresponding experimental value of 7-8~GPa~\cite{polian1990room} and hence results in an increase of $\kappa$ for $P>6$~GPa before the repulsive interactions between the atoms become so strong that compression of the structure is difficult and hence $\kappa$ decreases for $P>9$~GPa. 

The non-monotonic $P$-dependence of $\kappa$ is reflected in the structure, Fig.~\ref{Fig1}B, in that the peak positions and intensities of the radial distribution function $g_{\rm SiSi}(r)$ show a complex dependence on $P$, demonstrating that the network is modified in a highly non-affine manner. The Inset presents a cartoon relating the peak positions to the structural motifs of the network. Figure~\ref{Fig1}C shows that the position of the first peak, $r_1$, does indeed vary in a non-monotonic manner with $P$, in qualitative agreement with {\it ab initio} simulations~\cite{hasmy_percolation_2021}.
The initial decrease of $r_1$ (from $\approx 3.15$~\AA) with $P$ is due to the affine compaction of the tetrahedra (solid line), while the further reduction of  $r_1$ for $P>6$~GPa might be ascribed to a partial transformation of the structure to the metastable coesite-like phases~\cite{hu2015polymorphic, bykova2018metastable}, which are composed of a mixture of [SiO$_4$], [SiO$_5$] and [SiO$_6$] units, having $r_{\rm SiSi} \approx 2.95$~\AA. The increase of $r_1$ for $P > 9$~GPa might be related to a further structural transformation to Stishovite-like structures (all Si are six-fold coordinated and $r_{\rm SiSi} \approx 3.20$~\AA), likely to be triggered by the reduction of bond valency~\cite{okeeffe1981structure,hasmy_percolation_2021} and the formation of higher coordinated Si atoms at $P \approx 9$ GPa (see Fig.~\ref{Fig5}B).

The $P$-dependence of the second peak position, $r_2$, mirrors the one of $r_1$ since it is strongly affected by the nearest neighbor arrangement of the polyhedral units, Inset of panel b. 
That $r_2$ exhibits the strongest $P-$dependence of $r_i(P)$ implies that the modification of the polyhedral network on the scale of the rings is the primary densification mechanism, see also Fig.~\ref{SI_Fig4}. 
One thus concludes that the non-monotonic behavior of the change in $r_1$ and $r_2$ is directly related to the mechanical response of the system on the level of the rings. 
This non-monotonicity is strongly diminished for $r_3\approx 7.5$~\AA\ since this scale exceeds the typical diameter of a ring (Inset of Fig.~\ref{Fig1}B and Fig.~\ref{SI_Fig4}), making that $r_3$ decreases monotonically with $P$, although much slower than expected from an affine change of the structure, and the same is true for the peaks at larger distances.

For $P$ around 20~GPa, $r_1$ shows a plateau, indicating that at intermediate pressures the structure of glass cannot be compressed anymore on this length scale while the structure on larger scales is more flexible and hence can still be compressed. 
Beyond $P\approx 40$~GPa, the change of the peaks on various length scales exhibits the same linear dependence on $P$, i.e., the glass behaves like a linear-elastic solid. However, the fact that the decrease of the $r_i$ is significantly weaker than the one expected from an affine deformation shows that the structure is undergoing a complex compression process.


\begin{figure*}[ht]
\centering
\includegraphics[width=0.9\textwidth]{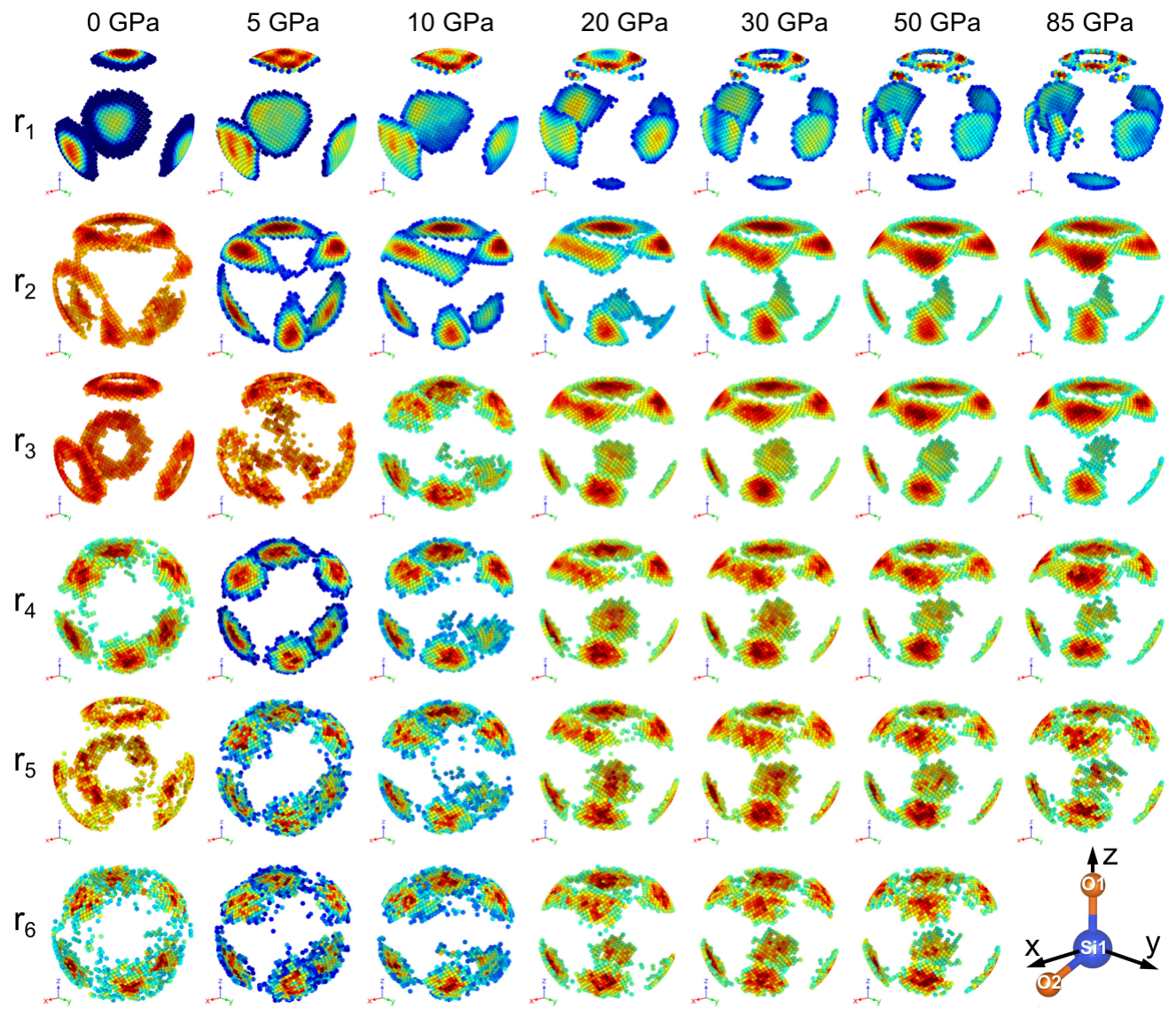}
\caption{ Three-dimensional structure as a function of distance and pressure.   
The density distribution of the Si atoms in a local reference frame formed by a Si and two neighboring O atoms (cartoon in the lower right corner). Only the area belonging to the top 25\% in density is shown and colored based on the normalized density which goes from 0.5 (dark blue) to 1 (dark red). 
} 1
\label{Fig2}
\end{figure*}

\bigskip
{\bf Many-body correlation and symmetry transitions.} To rationalize the complex $P$- and $r$-dependence of the structure seen in Fig.~\ref{Fig1}, we probe the structure in 3D. 
For this we introduce a local coordinate system based on a Si atom (Si1) and two of its nearest neighbor O atoms (O1 and O2)~\cite{zhang_revealing_2020}: Defining the position of Si1 as the origin, the direction from Si1 to O1 as the $z$-axis, and the plane containing the three particles as the $z$-$x$ plane (see schematic in Fig.~\ref{Fig2}), this local reference frame allows to introduce a spherical coordinate system $(r,\theta,\phi)$, and to measure the probability of finding any other atom at a given point in space, i.e., to measure a four-point correlation function.

\begin{figure}[ht]
\centering
    \includegraphics[width=0.4\textwidth]{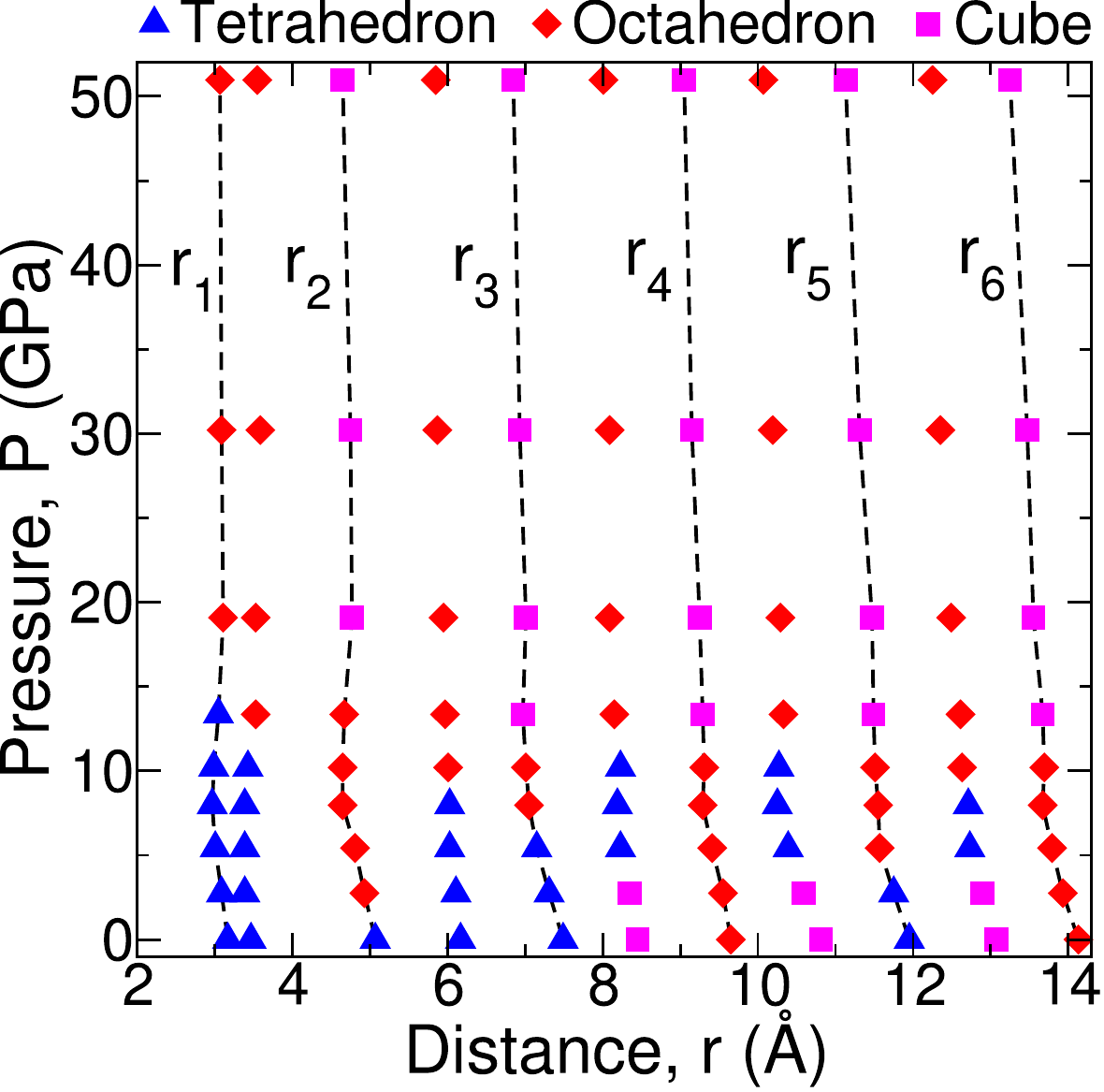}
\caption{Symmetry of the structure as a function of pressure and length scale. The shape of the symbols reflects the symmetry of the 3D Si-distribution at a given distance $r$. The points connected by the dashed lines correspond to the peak positions (marked by $r_n$) of $g_{\rm SiSi}(r)$, while the points between the lines correspond to the minima. 
} 
\label{Fig3}
\end{figure}

Figure~\ref{Fig2} shows that the resulting density field of Si atoms, $\rho(r,\theta,\phi)$, has for $P=0$ a pronounced tetrahedral symmetry for $r_1$, $r_3$ and $r_5$, while at $r_4$ and $r_6$ it is octahedral. 
The second shell features four rings, each of which is composed of three patches. The latter are created by the tetrahedral units that are in the second shell and are connected to a common tetrahedron in the first shell, see Fig.~\ref{SI_Fig5}. 
The octahedral symmetry at $r_4$ and $r_6$ is surprising since at this $P$ it does not reflect the symmetry of the central tetrahedron. Inspection of the particle arrangement shows that at this distance Si atoms start to occupy the directions of the center of the edges of the central tetrahedron, thus forming an octahedron, a structure that has higher local density than a tetrahedron, see Fig.~\ref{SI_Fig5}. Such an arrangement is only possible 
if the strands of connected Si-O bonds that emanate from the central Si atom can bend sufficiently
to access these directions, which is the case for $r_4$ and beyond, but not for smaller distances. It is thus this flexibility of the network on these intermediate length scales that permits to form particle arrangements that have a new symmetry.

With increasing $P$, the symmetry seen at $r_1$ changes from tetrahedron to decorated-tetrahedron ($P\gtrsim 10$~GPa), to decorated-octahedron ($P\gtrsim 20$~GPa). (These decorations are due to edge-sharing structures, see Fig.~\ref{SI_Fig6}.) By contrast, the symmetry seen at $r_2$ evolves from  a decorated-tetrahedron (see Fig.~\ref{SI_Fig5}) to an octahedron ($P=5-10$~GPa) and finally to a cube at high pressures ($P\gtrsim 20$~GPa).  
The $P$-dependence of $\rho(r,\theta,\phi)$ at $r_3$ resembles the one for $r_2$, except that the higher flexibility of the network on the scale of this larger distance permits the transitions to occurs at somewhat lower $P$s. For $r>r_1$ the symmetry of $\rho(r,\theta,\phi)$ becomes independent of $P$ if $\gtrsim 20$~GPa, demonstrating that the MRO becomes simple in terms of symmetry.

Figure~\ref{Fig3} presents the symmetry of $\rho(r,\theta,\phi)$ for distances that correspond to the maxima and minima in $g_{\rm SiSi}(r)$. (Figure~\ref{SI_Fig7} shows the density plots for the minima.) For small $P$ one finds at small and intermediate $r$ tetrahedra, while at larger distances there is an intricate dependence of the structure on $r$ due to the competition between achieving high packing density and the rigidity of the network structure on small length scales, a behavior that at somewhat higher $P$ is also found at short distances.  
For $P\gtrsim 10$~GPa one has instead a periodic alternation between cubes (at the maxima of $g_{\rm SiSi}(r)$) and octahedra (at the minima), since these two geometries are dual pairs and hence a low density area at one $r_i$ will correspond to a high density area in the following minimum and vice versa~\cite{zhang_revealing_2020}. (We note that the 3D density distribution of the O atoms is dual to the one of the Si atoms shown here, i.e., at a given $r$ the role of cube and octahedral symmetries are swapped.) 

In the past, the distribution of the coordination number has been used to estimate the entropy associated with the multiplicity of local motifs~\cite{wei2019assessing}.
 The density field $\rho(r,\theta,\phi)$ allows to make this idea more precise and to generalize it to quantify the number of possible local atomic arrangements at any distance $r$ by means of the Shannon entropy $H=-\sum_{i=1}^{K}{p_i}\ln p_i$. Here $K$ is the total number of pixels for the discretizations of the field, $p_i$ is the probability to find an atom in a specific direction of the local coordinate system, i.e., it is directly related to $\rho(r,\theta,\phi)$. The maximum value of $H$ corresponds to the uniform distribution $p_i=1/K$, giving $H_M=\ln(K)$.  

\begin{figure}[ht]  
\centering
\includegraphics[width=0.48\textwidth]{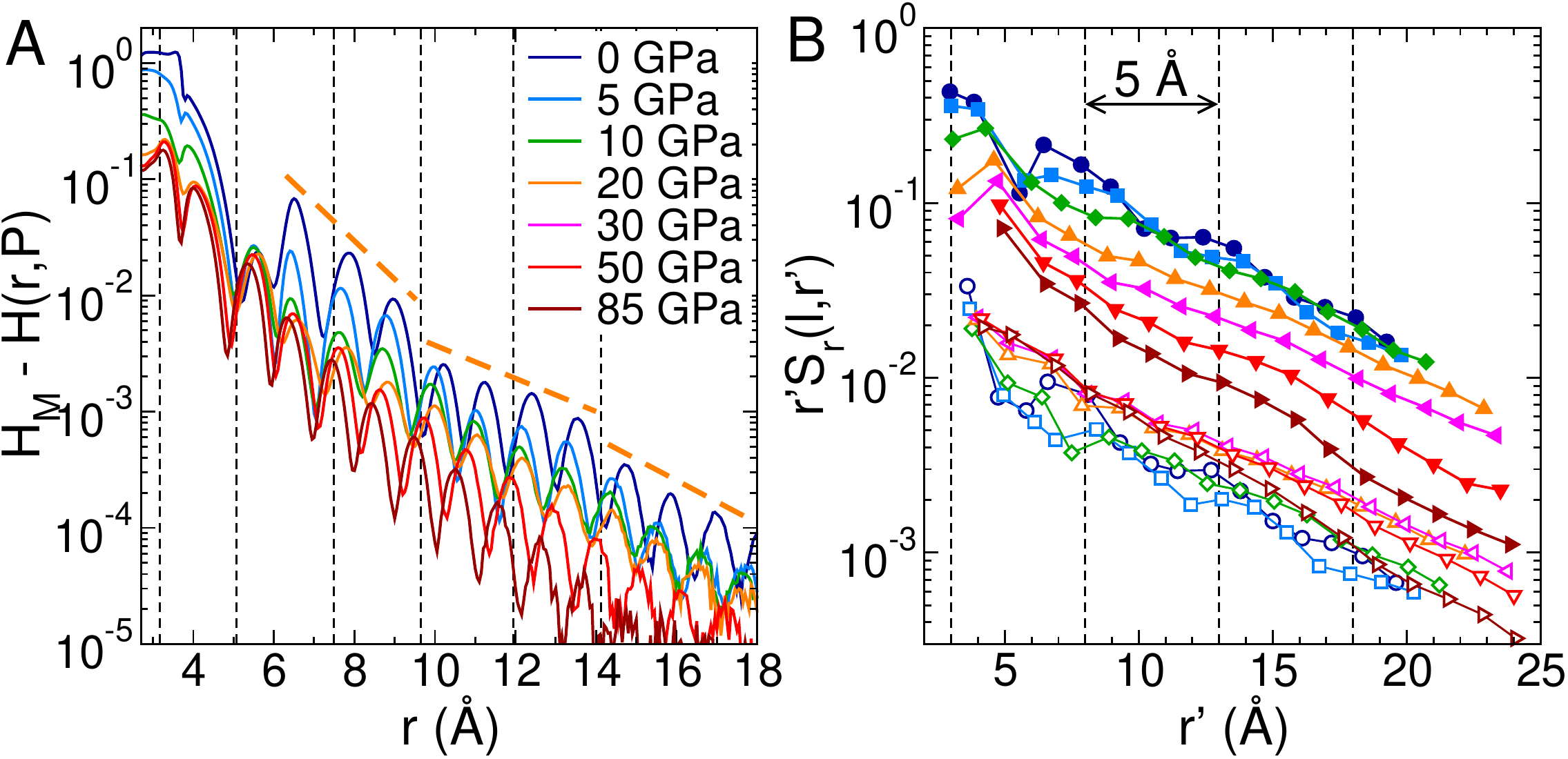}
\caption{Quantitative characterization of the three dimensional structure.
(A) The reduced Shannon entropy $H_M-H(r,P)$ measures the numbers of motifs in the structure on the length scale $r$. The vertical dashed lines indicate the positions of the maxima in $g_{\rm SiSi}(r)$ at $P=0$~GPa. 
The bold dashed lines highlight the change of local slope due to the presence of steps in the correlation function. 
(B) The correlation functions $r'S_\rho(3,r')$ (filled symbols) and $r'S_\rho(4,r')$ (open symbols), respectively. The curves $r'S_\rho(4,r')$ are divided by ten to avoid overlap with  $r'S_\rho(3,r')$. 
} 
\label{Fig4}
\end{figure}

\begin{figure*}[th]
\centering
\includegraphics[width=0.95\textwidth]{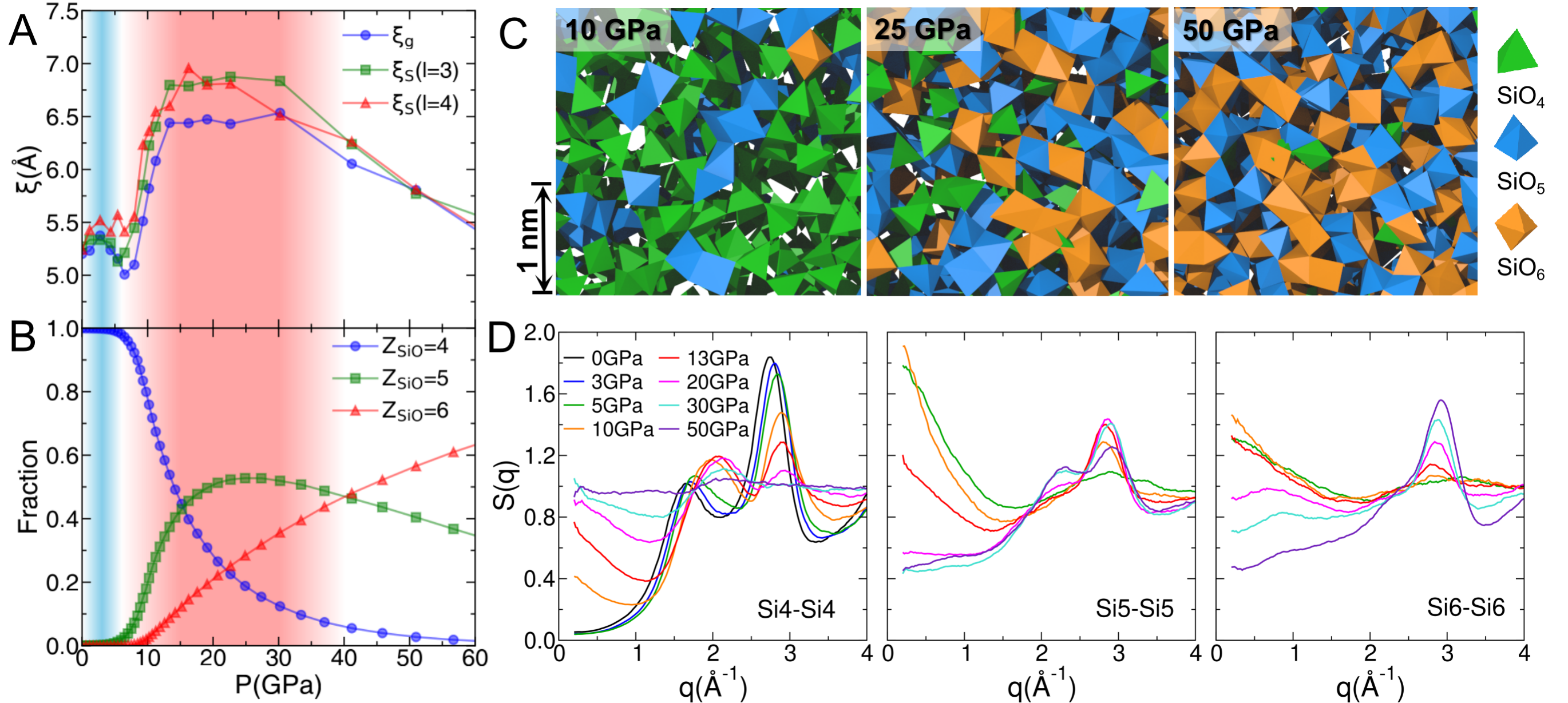}
\caption{Non-monotonic pressure-dependence of the medium-range order. (A) Structural correlation length $\xi$ as estimated from the decay of $S_\rho$ and $g(r)$. (B) Fraction of the Si species with a given coordination number. The bluish and the reddish background highlight the two maxima in $\xi$ at small and intermediate pressures, respectively. (C) Snapshots of the system at different pressures. The scale bar is 1~nm. The three relevant polyhedral units are shown in different colors. (D) Partial structure factor for Si atoms having 4, 5, and 6 nearest neighbor oxygen atoms (left to right). 
}   
\label{Fig5}
\end{figure*}

Figure~\ref{Fig4}A demonstrates that at small distances the reduced Shannon entropy $H_M-H(r,P)$ is large at $P=0$. 
 With increasing $P$ this difference drops quickly, indicating that the nearest neighbor shell becomes increasingly disordered due to compression. Independent of $P$, the signal has decayed by more than an order of magnitude at $r\approx 5$~\AA, reflecting the large number of structures on this length scale, thus rationalizing why silica is such a good glass-former. Surprisingly, one finds for low $P$s that the decay to zero at large $r$ occurs in steps (dashed lines), 
indicating the presence of a complex MRO. The width of these steps is around 5~\AA, corresponding to the ring-diameter that is most frequent at low $P$, see Fig.~\ref{SI_Fig4}, i.e., these steps indicate the presence of a packing order on the level of the rings. In contrast to this, the curves for high $P$ show no secondary modulation because the rings have collapsed and hence correspond no longer to a relevant length scale of the network. 
Apparently, the observed modulations in entropy are predominantly signatures of the network structure on the level of the rings. However, the change in network topology could have indirect consequences for the dynamics, e.g., by affecting stress/structure relaxation pathways~\cite{song2019atomic}.

The symmetry and strength of the orientational order can be quantified by expanding the density field in spherical harmonics and probing its power spectra $S_\rho(l,r)$, see Methods.  In agreement with earlier studies~\cite{zhang_revealing_2020,yuan2021prl,singh2023pnas,zhang2024prb}, these spectra show pronounced oscillations as a function of $r$, reflecting the changing symmetry of the structure as a function of the considered length scale, Fig.~\ref{SI_Fig8}. The observation of tetrahedral/cubic symmetries imply that the modes $l=3$ and $l=4$ are the relevant ones (see Figs.~\ref{SI_Fig8} and~\ref{SI_Fig9} for the $l$-and $P$-dependence of $S_\rho(l,r)$). 
One recognizes that
at low $P$ the signal for $l=3$ is significantly higher than the one for $l=4$, while at high $P$ the latter one dominates, reflecting the change of symmetry in the 3D particle arrangement, see Figs.~\ref{SI_Fig10} and~\ref{SI_Fig11}. 

The decay of the correlation function can be better understood by looking at the maxima of $r' S_\rho(3,r')$ as a function of $r'=L_0/L(P)$, a reduced distance that takes into account the affine part of the compression, Fig.~\ref{Fig4}B. (The multiplication of $S_\rho(3,r')$ by $r'$ is motivated by the Ornstein-Zernike relation~\cite{ornstein1914influence}.)  
These functions show broad peaks which occur at around 3~\AA, 8~\AA, 13~\AA, and 18~\AA, corresponding to the distances between a node in the network (Si atom) and the centers of the rings, Inset of Fig.~\ref{Fig1}B, i.e., the modulation of $S_\rho(3,r)$ reveals the symmetry of the structure on the scale of the rings, rationalizing the steps seen in the entropy.
With increasing $P$, this secondary modulation gradually disappears, showing that compression destroys the MRO induced by the rings. 
The $P$-dependence of the $l=4$ curves is relatively weak, Fig.~\ref{Fig4}B, implying that the strength of the octahedral/cubic symmetry is less sensitive to pressure than the tetrahedral symmetry. Surprisingly, the amplitude of the signal is non-monotonic in $P$, indicating that the MRO has a non-trivial $P$-dependence, and in the following we will reveal the origin of this effect.

\bigskip
{\bf Medium-range structural coherency.} At intermediate-to-large distances the correlation functions $r'S_\rho$ decay exponentially, $r'S_\rho \propto {\rm exp}(-r'/\xi_S)$, with the correlation length $\xi_S$,  and the same holds for $r'|g(r')-1|$, defining the length scale $\xi_g$, Fig.~\ref{Fig5}A.
One recognizes that the structural correlation lengths extracted from $g(r)$ and $S_\rho(l,r)$ are quantitatively similar. This can be rationalized by noting that $g(r)$ is closely related to the angular average of the four-point correlation function, and hence the two observables exhibit similar decay behavior~\cite{zhang_revealing_2020,zhang2024prb}.

Surprisingly, the $P$-dependence of $\xi_S$ and $\xi_g$ exhibits two maxima (see also Fig.~\ref{SI_Fig12}), with the first one at $P\approx 3$~GPa mirroring the compressibility maximum, see Fig.~\ref{Fig1}A. 
We note that although the structural correlation lengths characterize the MRO, their $P$-dependence is not always directly related to a change in the ring structure, as the typical characteristics of the rings, e.g., size and shape parameters (Fig.~\ref{SI_Fig4}) do not exhibit the same $P$-dependence as $\xi$. Instead, we rationalize the intriguing non-montonicity in $\xi$ as follows.
The first maximum at around 3~GPa can be rationalized by recognizing that a compression of the network, which at $P=0$ is under significant local stress (thus small $\kappa$), will result in a stress release ($\kappa$ grows), permitting a correlated elastic displacement of the atoms on a length scale that increases.
Further compression triggers reorientation of the tetrahedra that disrupts the structural order, thus $\xi$ decreases again. For $P \gtrsim 6$~GPa, plastic yielding sets in and the tetrahedra start to be replaced by other structural motifs, notably [SiO$_5$] and [SiO$_6$],  Fig.~\ref{Fig5}B. In particular, fivefold coordinated Si has been recognized as
an important species for the physical properties of silicate melts and glasses under pressure, notably for viscous flow and plastic deformation~\cite{stixrude2005structure,schulze2023fivefold,yuan2014brittle,liang2007mechanical}.
The presence of this species introduces additional Si-O bonds into the network, thus diminishing fluctuations in the local structure, which in turn results in an enhanced structural coherence of the network, i.e., $\xi$ increases. 
Indeed, Fig.~\ref{Fig5}A shows that the $P$-dependence of $\xi$ follows closely the concentration of [SiO$_5$], Fig.~\ref{Fig5}B, which first increases quickly up to around 20~GPa before starting to decrease at $P \gtrsim 30$~GPa. This decrease of the number of [SiO$_5$] entities, which are replaced by [SiO$_6$], leads to a decrease of $\xi$, since the latter have a packing structure that does not correspond anymore to an open network. 

The $P$-dependence of the structure can be directly seen in the snapshots of the system, Fig.~\ref{Fig5}C. On recognizes that at 10~GPa the random open network structure formed by [SiO$_4$] tetrahedra is locally modified by the [SiO$_5$] units. The latter structural motifs are not distributed randomly in the network but form fractal-like structures as becomes evident from the partial structure factor of the [SiO$_5$] units, Fig.~\ref{Fig5}D, which shows at low $q$ a significant increase. As a consequence of this local geometric change, also the structure factor of the [SiO$_4$] units shows an increase at low $q$, see the corresponding $S(q)$ in Fig.~\ref{Fig5}D. Further increase of the pressure to $P>20$~GPa, makes that the partial $S(q)$ for [SiO$_4$] becomes almost flat, i.e., these units are distributed basically randomly in the sample, while the $S(q)$ for [SiO$_5$] has no longer a peak at low $q$ but instead features a prominent peak at around 3~\AA$^{-1}$, i.e., these entities show a well organized structure like a normal dense liquid, rationalizing the relatively large correlation length at these pressures. If $P$ is increased to 50~GPa, one finds basically a random mixture of [SiO$_5$] and [SiO$_6$] units, see Fig.~\ref{Fig5}C. This impression is confirmed by the corresponding structure factors for these entities which show at this pressure a modest value at low $q$ and a substantial peak at around 3~\AA$^{-1}$. The disorder due to this mixing results in a small correlation length, in agreement with results for binary hard sphere mixtures for which it is found that high packing density is associated with a small $\xi$~\cite{yuan2021prl}.

\section*{Conclusions} 
The existence of a unit cell in crystals results in a symmetry of the structure that is independent of the length scale considered and one might expect that the disorder in glasses destroys this type of long-range order. We demonstrate for the prototypical network glassformer silica that this is not the case; instead, particles arrange in a manner that reflects at short distances the local motif of the structure, while at large scales one always finds an alternation between cubic and octahedral symmetries. 
Hence, although the evolution of local structure with pressure in silica glass bears some resemblance to crystalline silica polymorphs, this analogy should be understood only at the level of local structural units and their nearest-neighbor connections. At larger distances, the symmetry of the structure is primarily governed by the drive to reach a high packing fraction, which can be realized through cubic/octahedral particle arrangements. Such arrangements can form at intermediate-to-large distances even if the local structure does not have this symmetry, owing to the flexibility of the strands formed by connected Si-O bonds. At high pressures, the increased variability of local structures enhances this flexibility in the medium-range order, allowing cubic/octahedral symmetries to emerge already at shorter distances.

Since the relevant factors leading to this medium- and long-range structure are a common feature of all network forming glasses and melts, the identified particle arrangements—featuring alternating cubic and octahedral symmetries—are expected to be a generic characteristic of amorphous systems composed of tetrahedral building blocks, such as SiO$_2$, GeO$_2$, Si, Ge, and certain chalcogenides.
Our findings also open the door for obtaining deeper insights into material properties that depend on the medium range order, with implications ranging from the design of high-performance glass materials to the modeling of geological processes in the Earth's interior.

\section*{Methods}

\textbf{Simulation details.} Classical MD simulations for silica were carried out using the two-body effective potential proposed by Sundararaman \textit{et al.}~\cite{sundararaman_new_2018}, which has been found to give a good quantitative description of the density, structure and mechanical properties of silica. Its functional form is given by

\begin{equation} 
V(r_{\alpha\beta}) =  \frac{q_\alpha q_\beta e^2}{4\pi \epsilon_0 r_{\alpha\beta}} +
A_{\alpha\beta} \exp(-r_{\alpha\beta}/B_{\alpha\beta}) - \frac{C_{\alpha\beta}}{r_{\alpha\beta}^6} \quad , 
\label{eq:potential}
\end{equation}

\noindent
where $r_{\alpha\beta}$ is the distance between two atoms of species $\alpha$ and $\beta$, $q_\alpha$ is the value of the effective charge of $\alpha$, and $A_{\alpha\beta}$, $B_{\alpha\beta}$, $C_{\alpha\beta}$ are the potential parameters~\cite{sundararaman_new_2018}. In order to attain a high computational efficiency, the long-range interactions given by the Coulomb term in Eq.~(\ref{eq:potential}) are evaluated using the method proposed by Wolf {\it et al.}~\cite{wolf_exact_1999}. 
A previous study~\cite{zhang_potential_2020} has shown that this treatment of the long range forces has negligible influence on various properties of the system while allowing for a significantly improved computational efficiency, and thus large-scale simulations can be performed.

The silica sample contained 570,000 atoms, corresponding at $P=0$ to a cubic box size of 20~nm at the experimental glass density. Such a large system size is necessary to avoid noticeable finite size effects~\cite{zhang_potential_2020} and to attain stable results regarding the structural and mechanical responses of the sample under compression. The sample was first melted and maintained at 3000 K for 800 ps, a time span that is sufficiently long to equilibrate the liquids. 
Subsequently, the melt was cooled down to 300 K with a constant cooling rate of 0.25 K/ps. The as-produced glass sample was then annealed at 300 K for 160 ps. All these simulations were carried out in the isothermal–isobaric ($NPT$) ensemble at zero pressure.  

The glass at 300~K was then hydrostatically compressed up to 100~GPa, in a stepwise manner. Specifically, at each step the box dimensions were reduced by $0.005L_0$, where $L_0$ denotes the box dimensions at zero strain. Accordingly, the atoms are affinely displaced to match this deformation. Subsequently, the box was relaxed in the $NVT$ ensemble for 200~ps before the next deformation-relaxation cycle begins. This time span is long enough to permit the system to reach an annealed state that is not showing any drift in quantities like energy or pressure. This process is continued until the compressive stress reaches 100~GPa (corresponding to around 23\% of strain).
We emphasize that this compression is still in the limit of validity of the potential, i.e., one does not deviate too much from the typical bond lengths scales for which the potential has been tested~\cite{sundararaman_new_2018}. 
Figures~\ref{SI_Fig1}-\ref{SI_Fig3} show that the predicted $P$-dependence of the structure and mechanical properties (compressibility) agree well with experimental and {\it ab initio} simulation data, indicating that our simulations are able to reproduce the properties of real silica under compression.

Throughout the simulations, periodic boundary conditions were applied in all directions. Temperature and pressure were controlled using a Nos\'e-Hoover thermostat and barostat.
All simulations were carried out using the Large-scale Atomic/Molecular Massively Parallel Simulator software (LAMMPS)~\cite{plimpton_fast_1995} with a time step of 1.6~fs. 

\textbf{Ring structure.} We define the ring size $n$ as the number of Si atoms in the smallest closed loop of Si and O atoms, the so-called primitive rings~\cite{leRoux2010ring}, i.e., none of the identified rings can be decomposed into two smaller rings. The spatial extent of these rings can be characterized by their radius of gyration, $R_{\rm g}$, given by the expression
\begin{equation}
R_{\rm g}^2 = \frac{1}{2n}\sum_{i=1}^{2n} ({\vec{r}_i} - \vec{R}_{\rm C})^2 \quad,
\end{equation}
where $\vec{R}_{\rm C}$ is the location of the geometrical center of the ring. 
(Note that since a ring contains equal number of Si and O atoms, the total number of atoms constituting a $n$-membered ring is 2$n$.) 

To characterize the ring shape, we 
compute the geometrical inertia tensor of the ring defined as~\cite{kobayashi_machine_2023} 
 
\begin{equation}
M = \sum_{i=1}^{2n} \begin{pmatrix}
y_i^2+z_i^2 & -x_iy_i & -x_iz_i\\
-x_iy_i & x_i^2+z_i^2 & -y_iz_i\\
-x_iz_i & -y_iz_i & x_i^2+y_i^2
\end{pmatrix},
\end{equation}
where $x_i$, $y_i$, and $z_i$ are the coordinates of atom $i$ relative to the geometric center of the ring. 
Then, we determine the principal axes of this inertia tensor and define as $\vec{e}_k$ their normalized directions ($k=1,2,3$). The geometrical dimension of the ring is then defined via the length of a sides of the smallest cuboid that encloses the ring, i.e.,
\begin{equation}
l_k={\rm max}_i(\vec{e_k}\cdot \vec{r_i})-{\rm min}_i(\vec{e_k}\cdot \vec{r_i}).
\end{equation}

We define the ordering of $l_k$ such that $l_1<l_2<l_3$. Thus, the ratio $l_2/l_3$ indicates how elongated the ring is in the principal-plane.

\textbf{Four-point correlation functions.}
We construct a local coordinate system using a Si atom and any two of its nearest neighbor O atoms, {see the main text and the cartoon in Fig.~\ref{Fig2}. Note that this coordinate system can be defined for all triplets of neighboring particles, and these spatial density distributions can be averaged to improve the statistics.  Since this coordinate system is adapted to the local arrangement of the three particles, it allows to detect angular correlations in space that are not visible in $g(r)$ or other standard structural observables~\cite{zhang_revealing_2020}. 

The so-obtained spatial distribution of the particles, $\rho(\theta,\phi,r)$,  which is a four-point correlation function, can be analyzed in a quantitative manner by decomposing it into spherical harmonics $Y_l^m$,

\begin{equation}
\rho(\theta,\phi,r) = \sum_{l=0}^\infty \sum_{m=-l}^{l}\rho_l^m(r) Y_l^m(\theta,\phi),
\label{eq7}
\end{equation}

\noindent
where the expansion coefficients $\rho_l^m$ are given by

\begin{equation}
\rho_l^m (r)=\int_0^{2\pi} d\phi \int_0^\pi d\theta \sin \theta \rho(\theta,\phi,r) Y_l^{m*}(\theta,\phi) \quad.
\label{eq8}
\end{equation}

Here $Y_l^{m*}$ is the complex conjugate of the spherical harmonic function of degree $l$ and order $m$.  In practice, the integral is carried out by sampling the integrand at a given radius $r$ over up to $10^8$ points using a shell of width 1.0~\AA. The strength of the 3D order can then be characterized by the square root of the angular power spectrum,  

\begin{equation} 
S_\rho(l,r)=\Bigl[(2l+1)^{-1}\sum_{m=-l}^{l}|\rho_l^m(r)|^2\Bigr]^{1/2}.
\label{eq9}
\end{equation}

We measured the structural correlation length $\xi_g$ by fitting $r'|g(r')-1|$ for $r > 6$ \AA, whereas $\xi_S$ was estimated by fitting $r'S_\rho$ for $r>7$~\AA. We found that the measured correlation lengths are stable with respect to a moderate change of the fitting ranges.

\section*{Data Availability}
The data that support the findings of this study are available upon reasonable request from the corresponding authors.

 \section*{Acknowledgements}
We thank S.~K.~Lee, K.~U.~Hess, and H.~Tanaka for discussion. The work was supported by the National Natural Science Foundation of China (Grant No. 12474185). W.K. acknowledges support from the Institut universitaire de France.

\section*{Author contributions}
W.K. and Z.Z. conceived the project. Z.Z. performed the MD simulations. All authors contributed to data analysis. Z.Z. and W.K. wrote the paper with input from Z.X.

\bibliography{refs}

\clearpage

\begin{widetext} 
\centering
\noindent \Large {\bf Supplementary Information for ``Symmetry Transitions Beyond the Nanoscale in Pressurized Silica Glass"}
\\
\large
\noindent Zhen Zhang$^1$, Zhencheng Xie$^1$, Walter Kob$^2$

{\it
\noindent $^1$College of Physics, Chengdu University of Technology, Chengdu 610059, China
\\
\noindent $^2$Department of Physics, University of Montpellier, CNRS, F-34095 Montpellier, France
}

\vspace{10mm}

\pagenumbering{arabic}
\setcounter{page}{1}

\renewcommand{\figurename}{Figure}
\renewcommand{\thefigure}{S\arabic{figure}}
\setcounter{figure}{0}

\begin{figure*}[ht]
\centering
\includegraphics[width=0.8\textwidth]{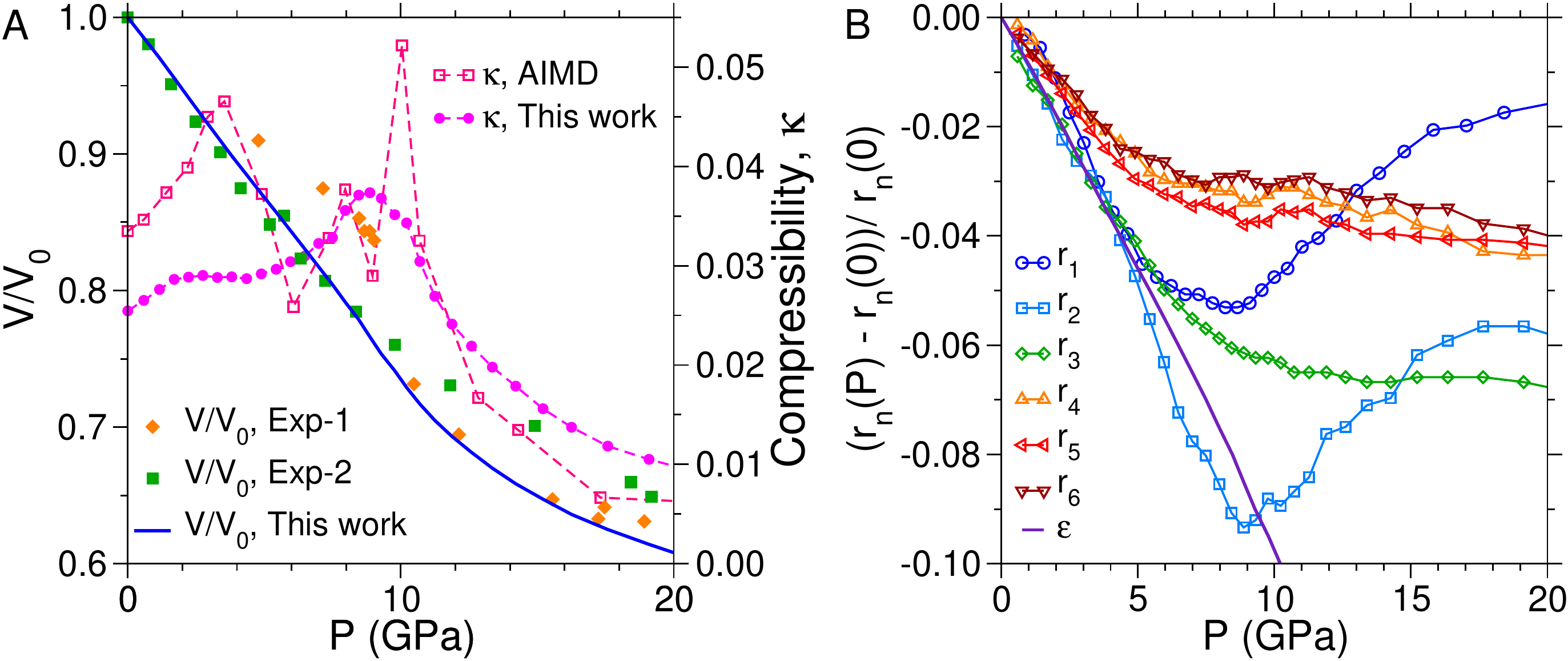}
\caption{(A) Pressure dependence of the reduced volume and compressibility for $P=0-20$~GPa. The experimental data for $V/V_0$ are from Refs.~[1,2]. 
 The {\it ab initio} molecular dynamics (AIMD) data for $\kappa$ are from Hasmy et al.~[3]. 
 One sees that both the density and compressibility match well with the reference data. (B) The relative change of the peak positions for $P=0-20$~GPa (zoom of Fig.~\ref{Fig1}B in the main text).
}
\label{SI_Fig1}
\end{figure*}

\begin{figure*}[ht]
\center
    \includegraphics[width=0.9\textwidth]{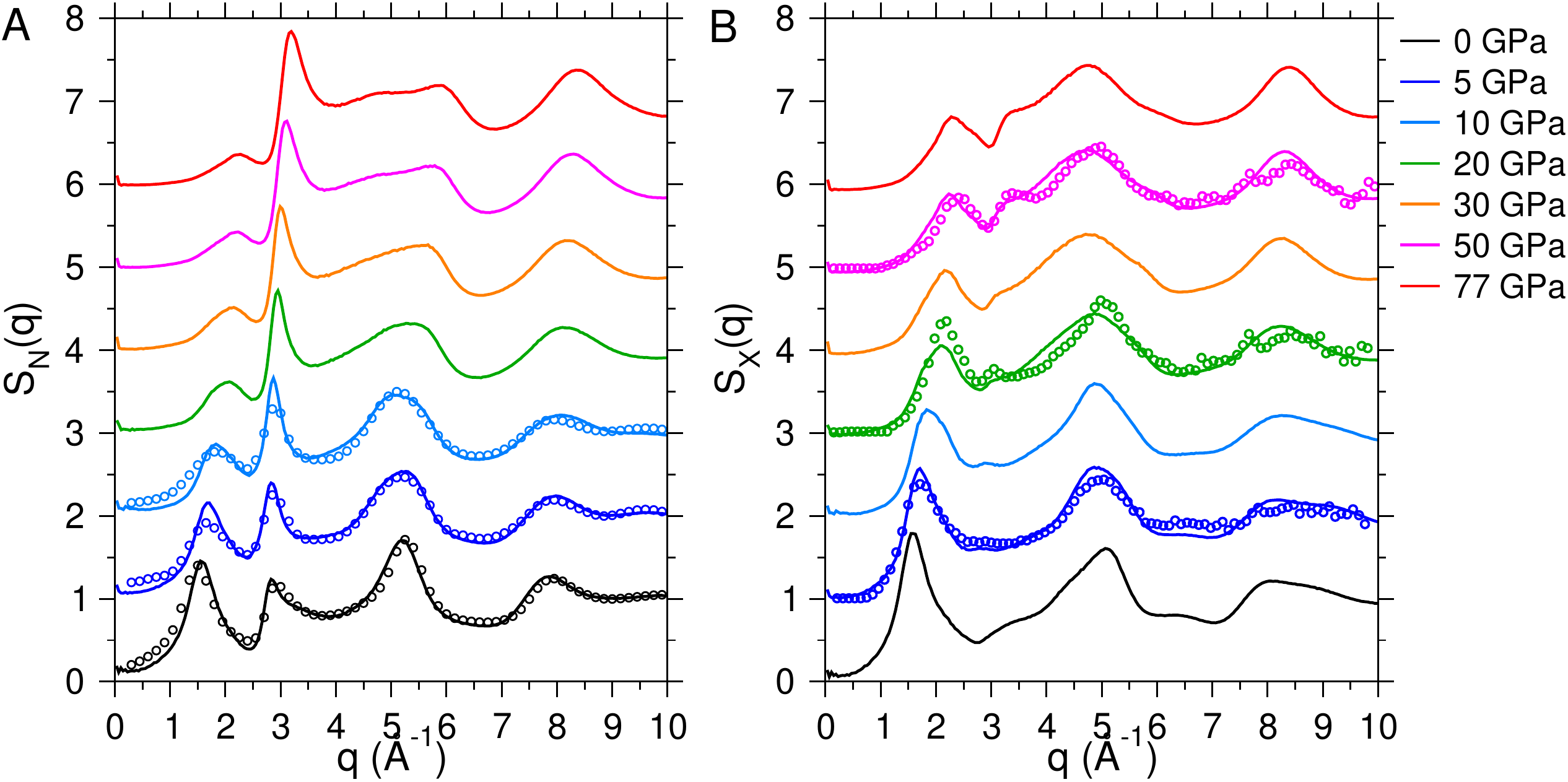}
    \caption{Simulated neutron structure factor $S_N(q)$, panel (A), and X-ray structure factor $S_X(q)$, panel (B), compared with the experimental data (open symbols) by Zeidler et al.~[4] (neutron) and Prescher et al.~[5] (X-ray). These functions have been obtained from the partial structure factors determined from the atomic configurations. The simulation results agree well with experiments over a wide pressure range. 
    }
    \label{SI_Fig2}
\end{figure*}

\begin{figure*}[ht]
\center
\includegraphics[width=0.5\textwidth]{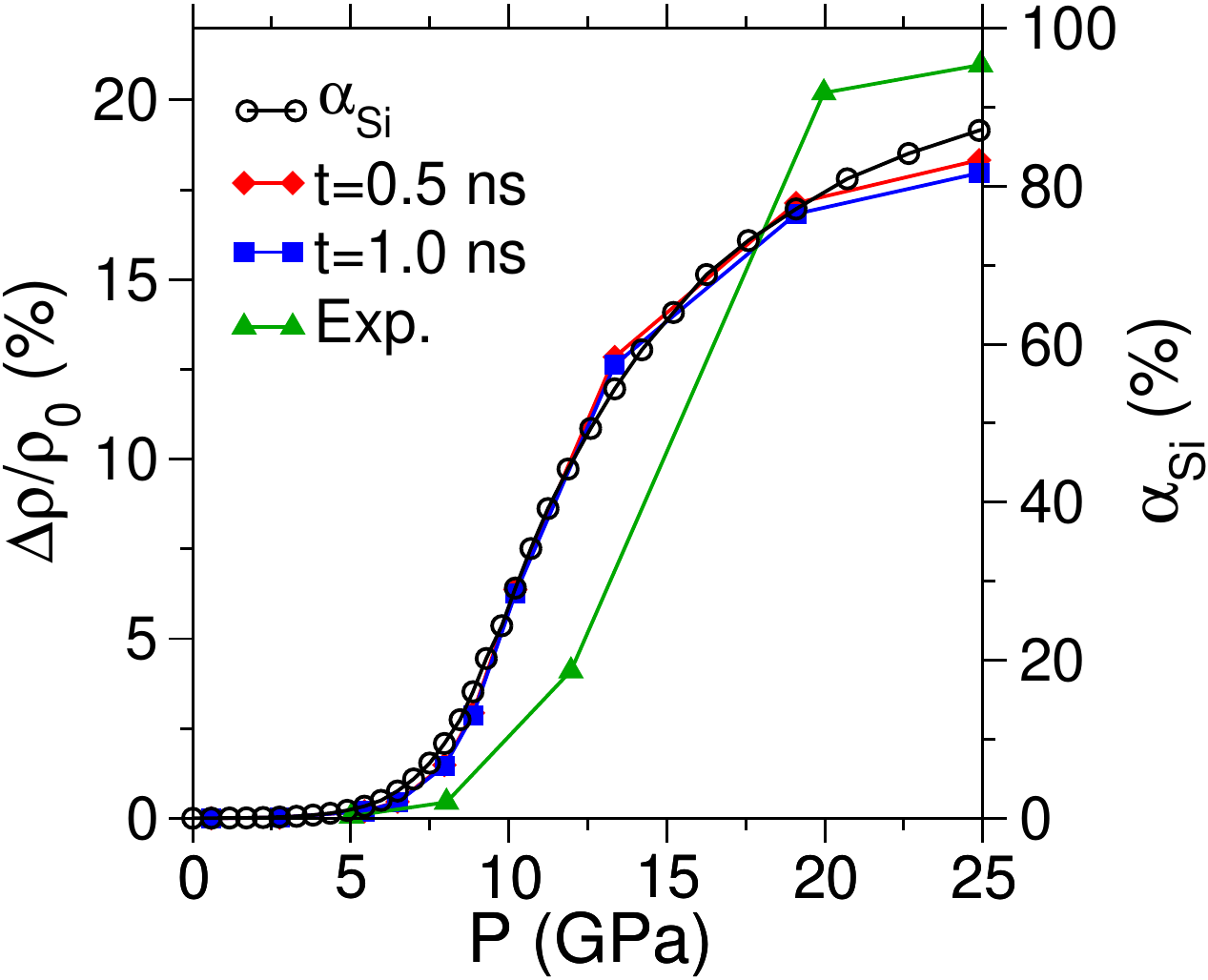}
\caption{Left ordinate: 
Relative difference in density between the as quenched glass and the sample that has been compressed to pressure $P$, kept for a time $t$ at this pressure (see legend for values of $t$), and then uncompressed to $P=0$. Experimental data are from~[6].
Right ordinate: Fraction of Si atoms that have,  with respect to the initial undeformed configuration, changed their bonding environment when compressed to pressure $P$. This allows to see to what extent the Si-O network topology changes during compression. 
}
\label{SI_Fig3}
\end{figure*}

\begin{figure*}[ht]
\center
\includegraphics[width=0.9\textwidth]{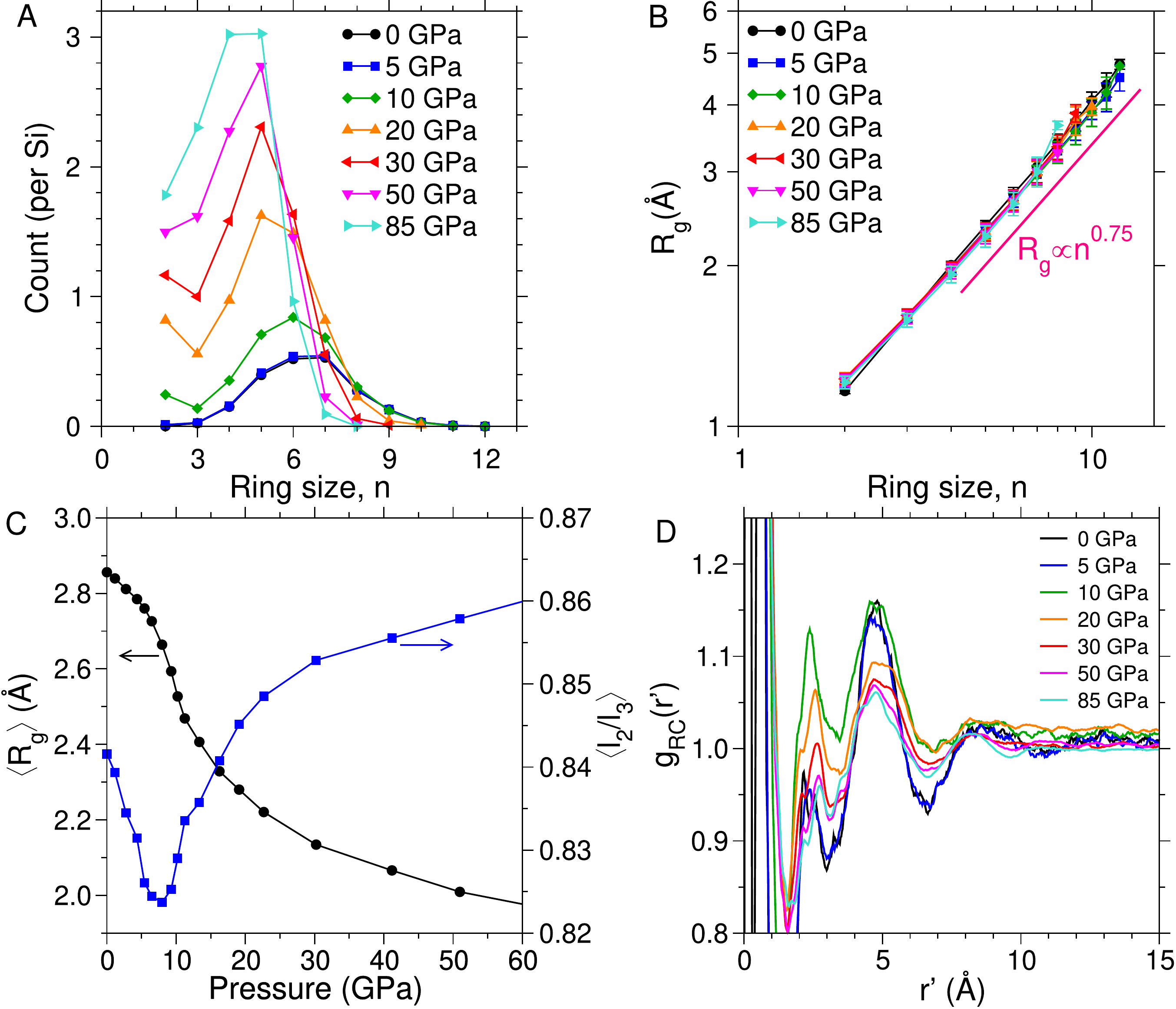}
\caption{Properties of the rings. (A) Ring statistics. The normalization is done with respect to the number of Si atoms. One sees that the number of rings increases with pressure. Note that the curves for $P=0$~GPa and $P=5$~GPa basically coincide. (B) Power-law dependence of the radius of gyration $R_{\rm g}$ of a ring as a function of its size $n$. The exponent is around 0.75, independent of the pressure. (C) Left ordinate: The average ring size $\langle R_{\rm g} \rangle$ vs. pressure. Right ordinate: The aspect ratio $l_2/l_3$ quantifies how much the shape of a ring deviates from a circle.  
(D) The radial distribution function $g(r)$ for the ring centers (RC). 
For small $P$, the ring size distribution remains unchanged, panel A. However, the rings become more elongated, panel C, indicating that densification of the sample results in a squashing of the rings instead of their compression in an isotropic manner, i.e., the prediction of simple elastic theory~[7]. Hence, at low $P$ the rings are squashed to close the voids without significant change of the network topology. Further compression beyond the elastic region results in the transformation of large rings to smaller rings, panel A, and the newly formed rings are more rounded than the collapsed large rings, resulting that the mean aspect ratio of the rings increases, panel C. 
}
    \label{SI_Fig4}
\end{figure*}

\begin{figure*}[ht]
\center
\includegraphics[width=0.9\textwidth]{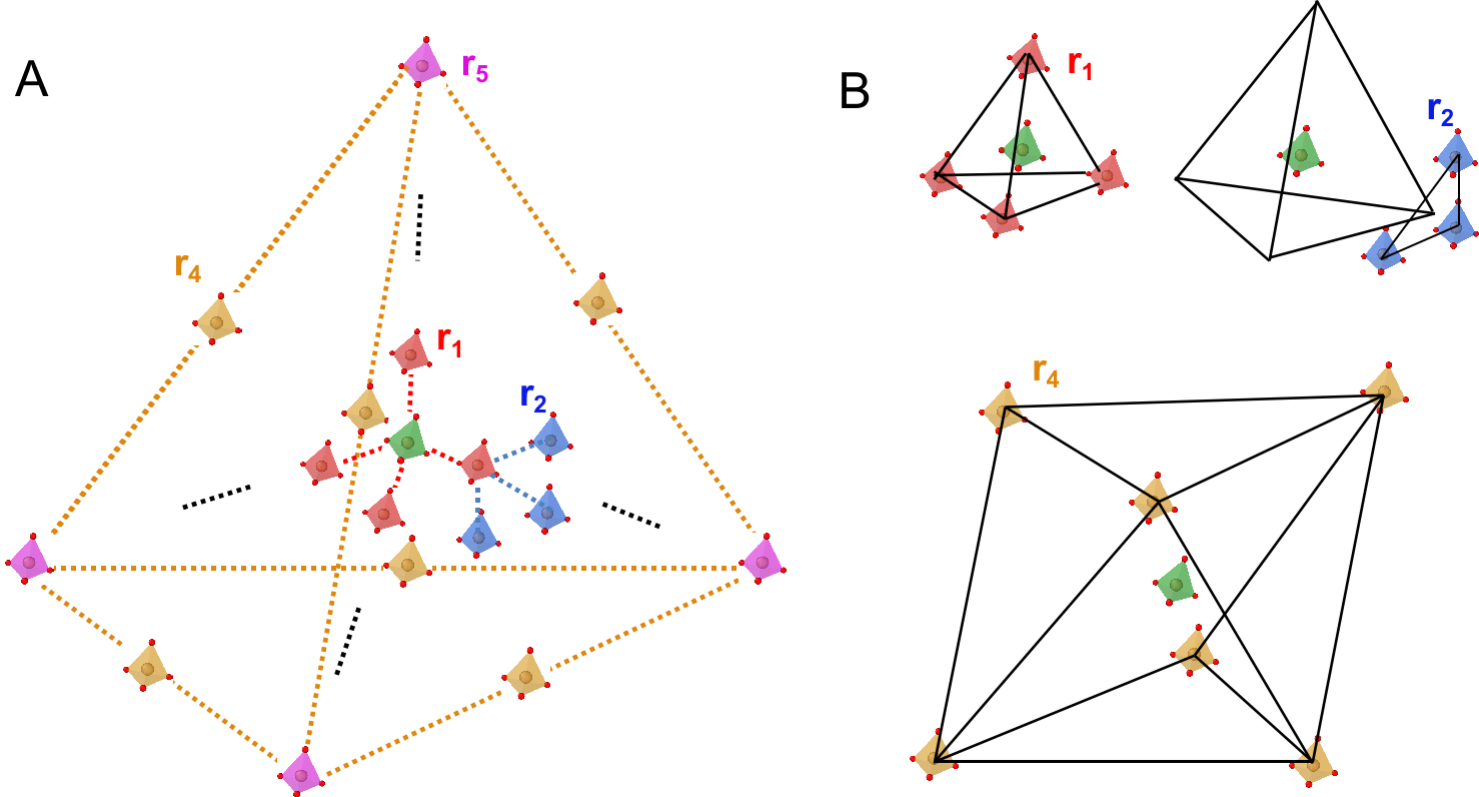}
\caption{(A) Schematic representation of the spatial arrangement of the Si atoms. The tetrahedral units they belong to are also shown and colored based on the distance (labeled $r_n$) from the central Si atom (green). 
For the sake of clarity, the second shell only shows the three tetrahedra connected to a node of the first shell, and the atomic configuration is enlarged in an affine way, making that the Si atoms are more distant from each other to allow better visualizing the geometry. 
For $r_4$, one finds an octahedral symmetry since the sequence of four Si-O links is sufficiently flexible that the forth Si atom can occupy the direction of the edges of the central tetrahedron.
(B) The first, second, and fourth shells are presented with their symmetries highlighted by the solid lines.
Note that beyond the nanoscale ($r\gtrsim r_4$) the atomic configuration giving rise to a high-density spot in Fig.~\ref{Fig2} is not a single tetrahedral unit, but an intermediate-sized clusters (super-structures) formed by nearest neighbor polyhedral units.
}
\label{SI_Fig5}
\end{figure*}

\begin{figure*}[ht]
\center
\includegraphics[width=0.9\textwidth]{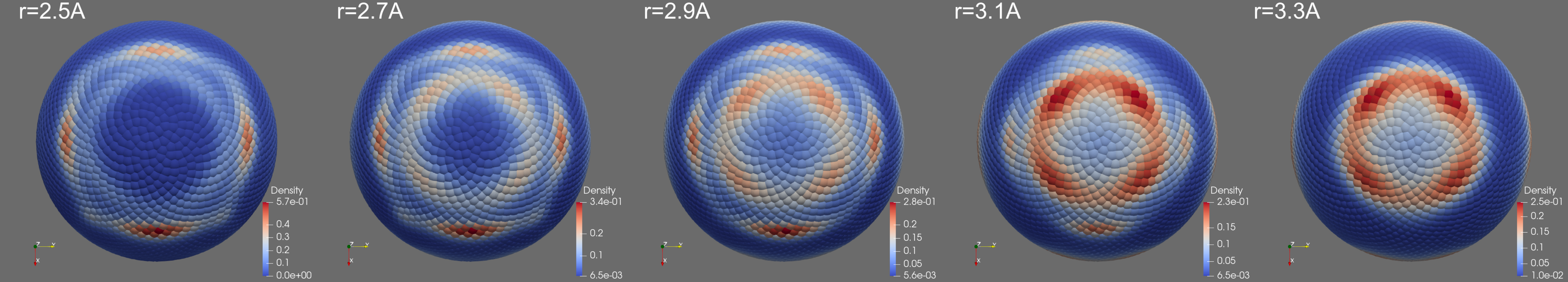}
\caption{Top view of the density field of the SiSi correlation for $P=50$~GPa. The graph for $r=3.1$~\AA\ corresponds roughly to the snapshot at $r_1$ and $P=50$~GPa in Fig.~\ref{Fig2}. The shown distances span the two peaks seen in $g_{\rm SiSi}(r)$ (Fig.~\ref{Fig1}B); the first peak at $r\approx2.5$~\AA\ arises from edge-sharing polyhedral connections and the main peak at $r\approx3.1$~\AA\ is due to the corner-sharing polyhedral connections. The ``decorations'' that one sees in the snapshot at $r_1$, i.e., the 4 halos seen just outside the main ring are due to the contribution to $\rho(r,\theta,\phi)$ from the edge-sharing structures. 
}
    \label{SI_Fig6}
\end{figure*}

\begin{figure*}[ht]
\centering
\includegraphics[width=0.9\textwidth]{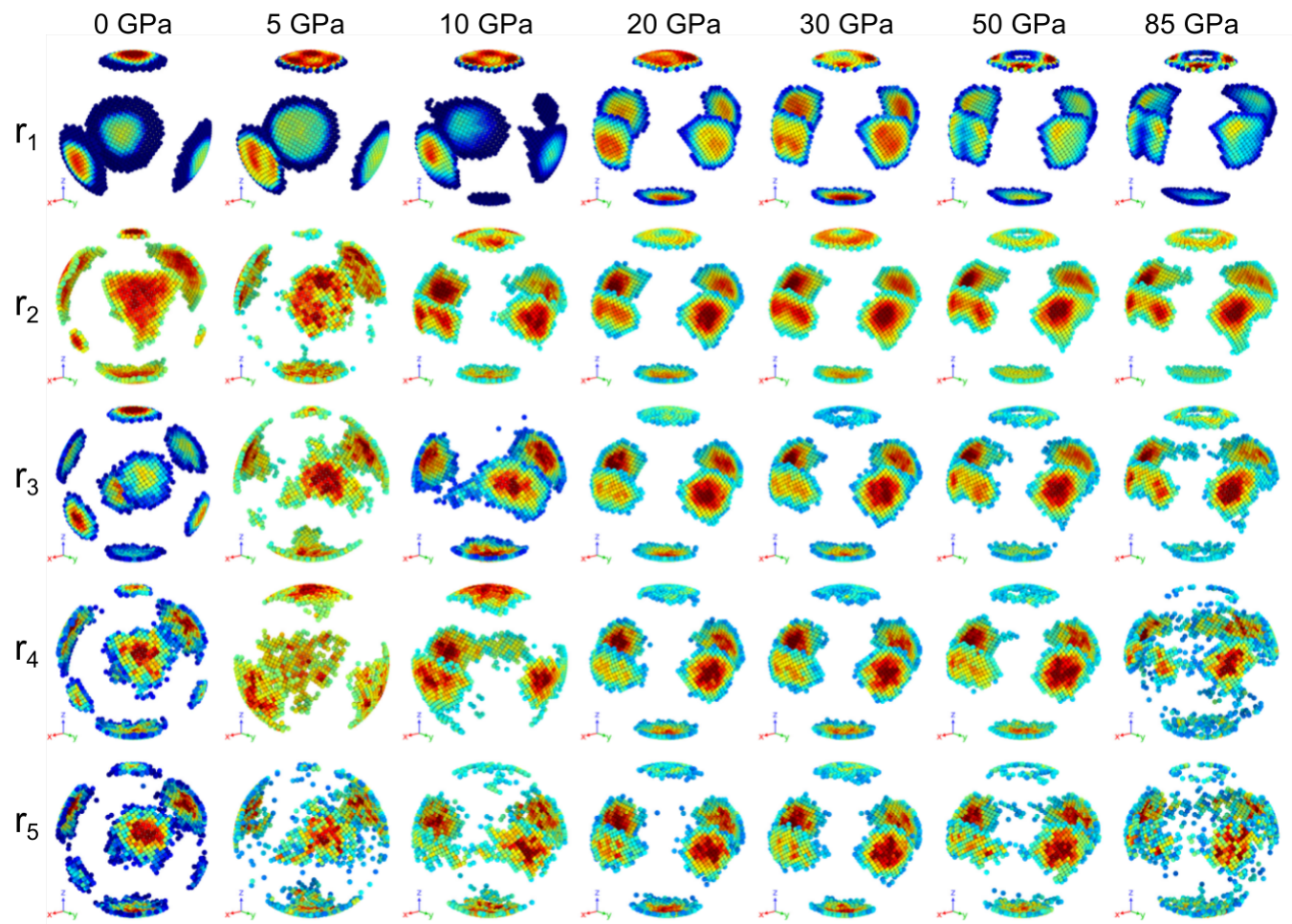}
\caption{Density plots for the distances corresponding to the minima of $g_{\rm SiSi}(r)$. Only the top 25\% of cells are shown and colored based on the normalized density which goes from 0.5 (dark blue) to 1 (dark red). 
} 
\label{SI_Fig7}
\end{figure*}

\begin{figure*}[ht]
\center
\includegraphics[width=\textwidth]{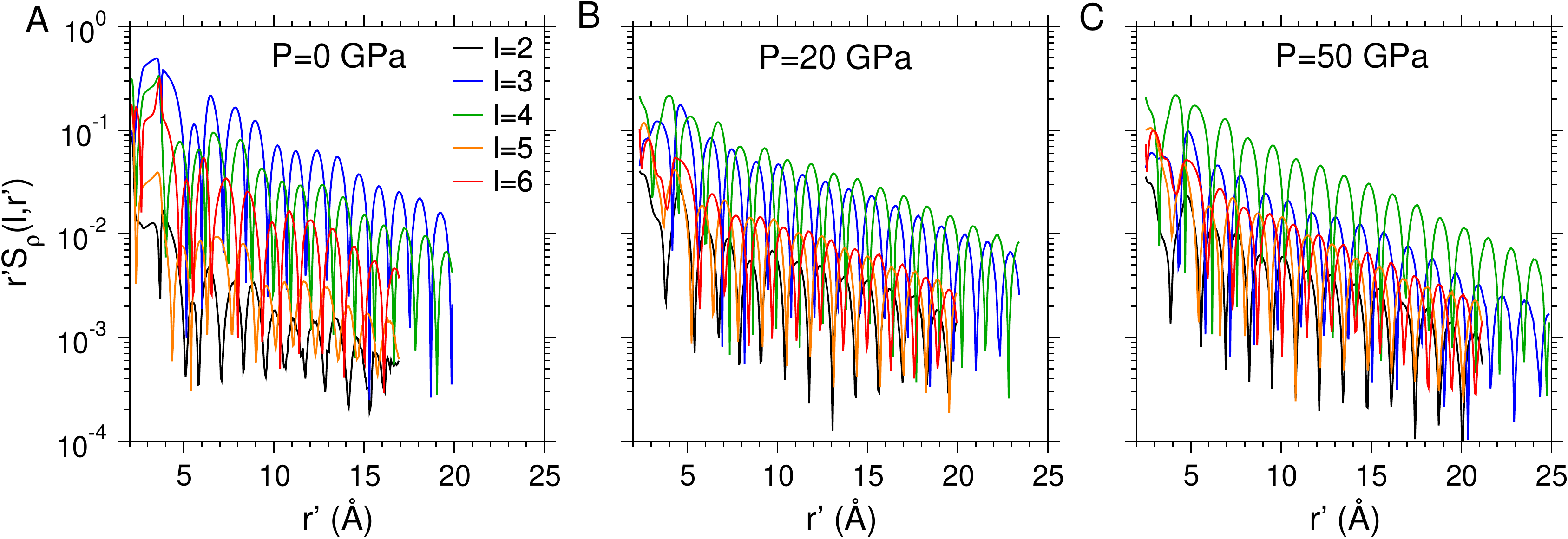}
\caption{$l$-dependence of $r'S_\rho(l,r')$, where the scaled distance is given by $r'=rL_0/L(P)$, with $L_0$ and $L(P)$ the box lengths at zero pressure and pressure $P$, respectively. The modes $l=3$ and $l=4$ are the most pronounced ones at low and high pressures, respectively. 
}
\label{SI_Fig8}
\end{figure*}

\begin{figure*}[ht]
\center
\includegraphics[width=0.4\textwidth]{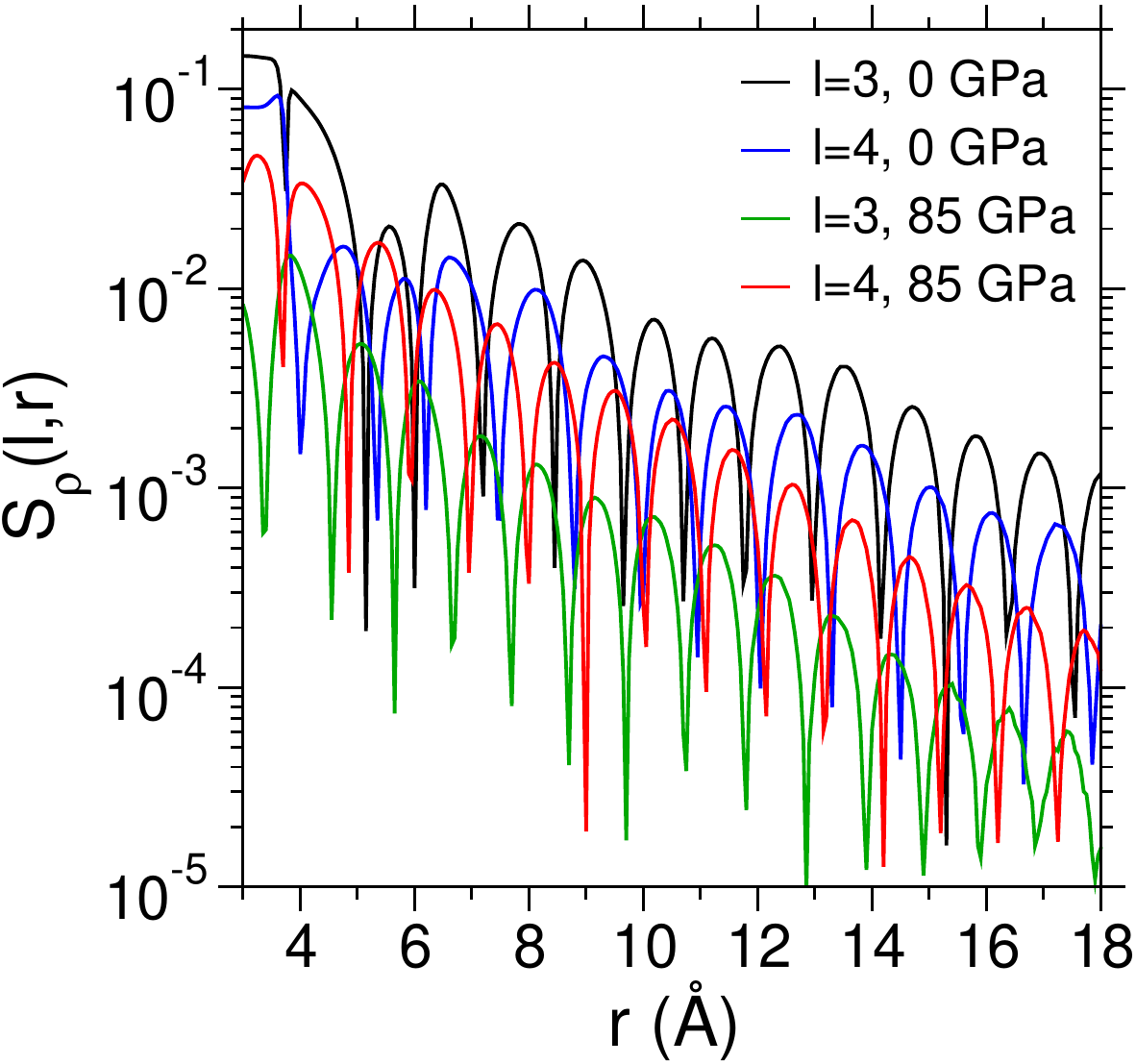}
\caption{The $r$-dependence of the power spectra $S_\rho(l,r)$ for $l=3$ and $l=4$.}
\label{SI_Fig9}
\end{figure*}

\begin{figure*}[ht]
\centering
\includegraphics[width=0.8\textwidth]{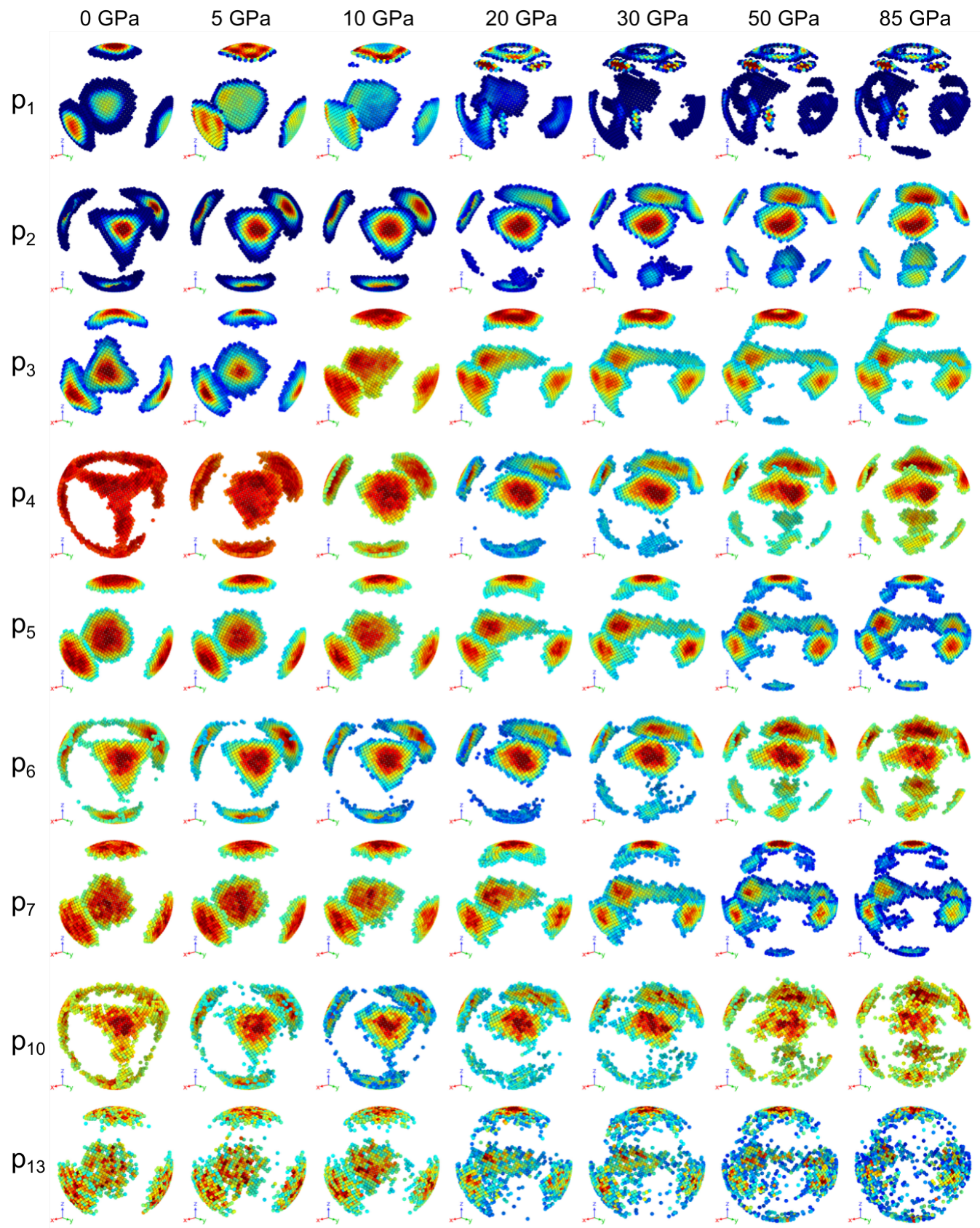}
\caption{Density plots for the distances corresponding to the maxima of $S_\rho(3,r)$. $p_n$ corresponds to the $n$th peak of $S_\rho(3,r)$. For $P=0$~GPa, the 1st and 13th peaks are at $r\approx 3$~\AA\ and 17~\AA, respectively, see Fig.~\ref{SI_Fig8}. For $l=3$ at $P\leq 10$~GPa, one observes that a highly regular alternating sequence of tetrahedral/anti-tetrahedral symmetry persists up to large distances. With increasing $P$, the local tetrahedral symmetry is destroyed and the alternating symmetries gradually make a transition to 
octahedron/cube. 
} 
\label{SI_Fig10}
\end{figure*}

\begin{figure*}[ht]
\centering
\includegraphics[width=0.85\textwidth]{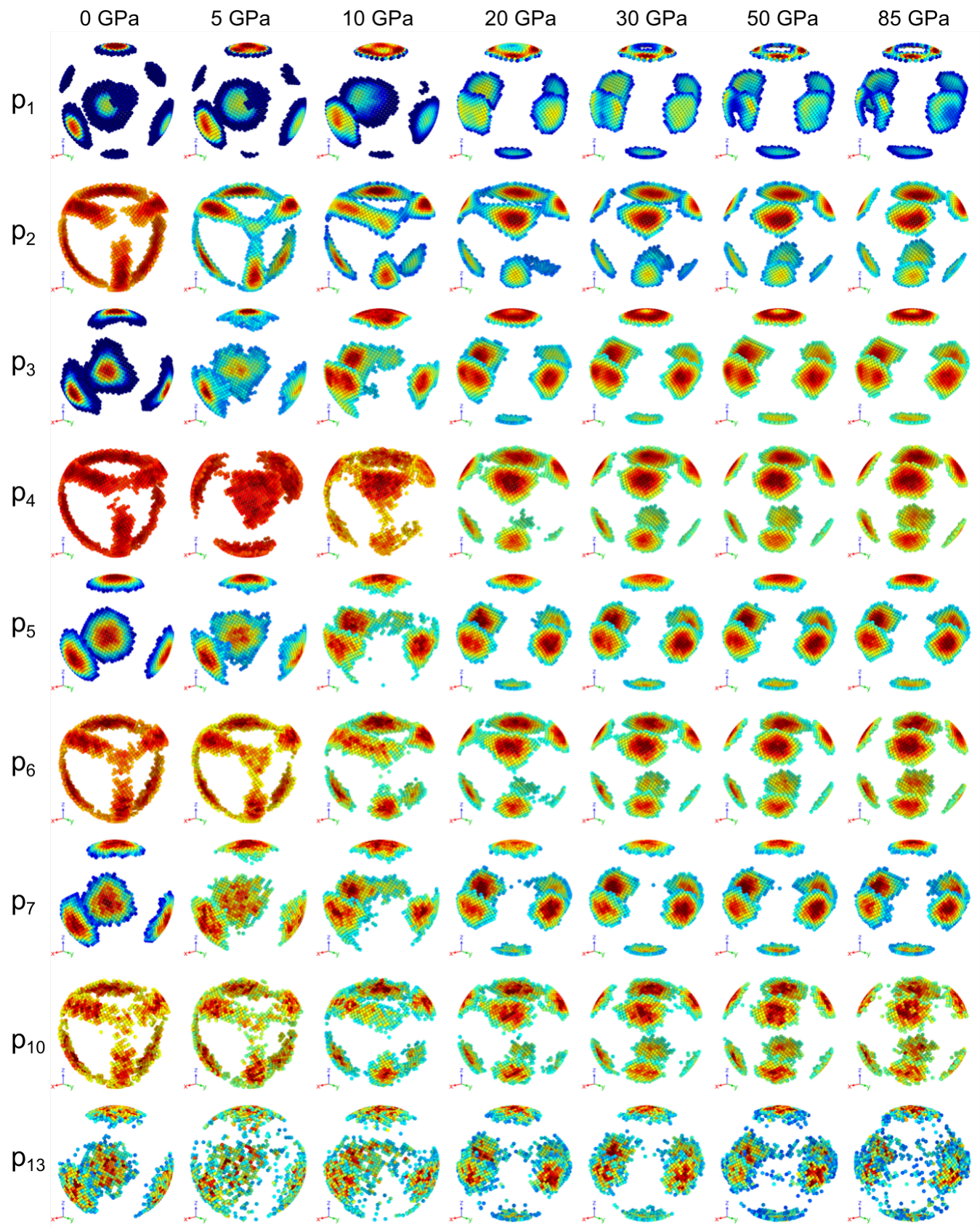}
\caption{Density plots for the distances corresponding to the maxima of $S_\rho(4,r)$. $p_n$ corresponds to the $n$th peak of $S_\rho(4,r)$. For $P=0$~GPa, the 1st and 13th peaks are at $r\approx 3.6$~\AA\ and 17.3~\AA, respectively, see Fig.~\ref{SI_Fig8}. The $P$-dependence of $l=4$ is similar to the one of $l=3$, but the former mode represents better the octahedral/cubic symmetries. 
} 
\label{SI_Fig11}
\end{figure*}

\begin{figure*}[ht]
\center
\includegraphics[width=0.85\textwidth]{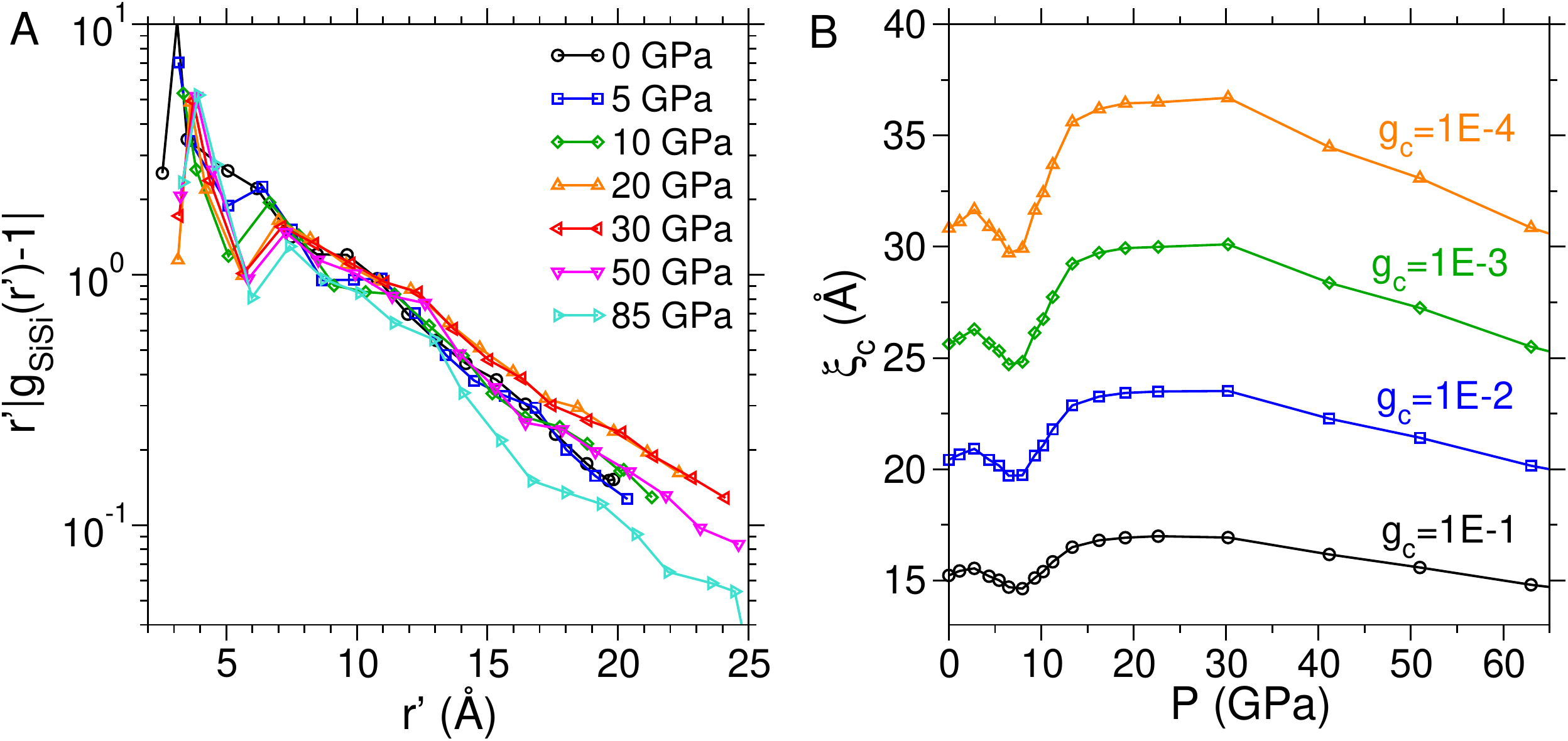}
\caption{(A) The maxima of $r'|g_{\rm SiSi}(r')-1|$. (B) Structural correlation length $\xi_c$ defined as the distance where the correlation function $r'|g(r')-1|$ decays to a predefined threshold value $g_c$. One observes that the magnitude of $\xi_c$ increases with decreasing $g_c$ but the non-monotonic behaviors, at both small-$P$ and large-$P$ regions, are the same.
}
\label{SI_Fig12}
\end{figure*}

\clearpage
\section*{References}
\raggedright \normalsize [1] H. Sugiura, K. Kondo, and A. Sawaoka, Dynamic response of fused quartz in the permanent densification region, J. Appl. Phys. 52, 3375 (1981).\\
\raggedright \normalsize [2] T. Sato and N. Funamori, High-pressure structural transformation of SiO2 glass up to 100
GPa, Phys. Rev. B 82, 184102 (2010).\\
\raggedright \normalsize [3] A. Hasmy, S. Ispas, and B. Hehlen, Percolation transitions in compressed SiO2 glasses, Nature
599, 62 (2021).\\
\raggedright \normalsize [4] A. Zeidler, K. Wezka, R. F. Rowlands, D. A. Whittaker, P. S. Salmon, A. Polidori, J. W.
Drewitt, S. Klotz, H. E. Fischer, M. C. Wilding, C. L. Bull, M. G. Tucker, and M. Wilson, High-Pressure Transformation of SiO2 Glass from a Tetrahedral to an Octahedral Network:
A Joint Approach Using Neutron Diffraction and Molecular Dynamics, Phys. Rev. Lett. 113,
135501 (2014).\\
\raggedright \normalsize [5] C. Prescher, V. B. Prakapenka, J. Stefanski, S. Jahn, L. B. Skinner, and Y. Wang, Beyond
sixfold coordinated Si in SiO2 glass at ultrahigh pressures, Proc. Natl. Acad. Sci. U.S.A. 114,
10041 (2017).\\
\raggedright \normalsize [6] T. Rouxel, H. Ji, T. Hammouda, and A. Moréac, Poisson’s ratio and the densification of glass
under high pressure, Phys. Rev. Lett. 100, 225501 (2008).\\
\raggedright \normalsize [7] L. D. Landau, L. Pitaevskii, A. M. Kosevich, and E. M. Lifshitz, Theory of elasticity: 3rd
Edition (Butterworth-Heinemann, 1986)

\end{widetext}
\end{document}